\documentclass[12pt]{article}
\InputIfFileExists{epsf.sty}{}{}
\usepackage{graphicx}
\def\beq{\begin{equation}}
\def\eeq{\end{equation}}
\def\bea{\begin{eqnarray}}
\def\eea{\end{eqnarray}}
\def\pew{P_{EW}}
\def\pewc{P^C_{EW}}
\def\expg{e^{i \gamma}}
\def\nn{\nonumber}
\def\sss{\scriptscriptstyle}
\def\roughly#1{\mathrel{\raise.3ex\hbox
{$#1$\kern-.75em\lower1ex\hbox{$\sim$}}}}

\def\gsim{\roughly>}

\begin{document}

\begin{flushright}  
UdeM-GPP-TH-04-124\\
\end{flushright}

\begin{center}
\bigskip
{\Large \bf CP Violation in $B\to \pi K$ Decays \footnote{talk given at {\it MRST 2004: From Quarks to
Cosmology}, Concordia University, Montr\'eal, May 2004.}} \\
\bigskip
\bigskip
{\large Maxime Imbeault \footnote{maxime.imbeault@umontreal.ca}}
\end{center}

\begin{center}
{\it Laboratoire Ren\'e J.-A. L\'evesque, Universit\'e de Montr\'eal,}\\
{\it C.P. 6128, succ. centre-ville, Montr\'eal, QC,
Canada H3C 3J7} 
\end{center}

\begin{center} 
\bigskip (\today)
\vskip0.5cm
{\Large Abstract\\}
\vskip3truemm
\parbox[t]{\textwidth} {I briefly review CP violation in the $B$ system,
concentrating on $B\to \pi K$ decays. I discuss how to deal with
electroweak-penguin contributions to these decays using flavour
SU(3). With these, I show that the entire unitarity triangle can be
extracted from measurements of $B\to \pi K$ decays. Finally, I examine
the signals for new physics in these decays and the possibilities for
measuring them.}
\end{center}

\thispagestyle{empty}
\newpage
\setcounter{page}{1}
\baselineskip=14pt

\section{Introduction} 

There are many known methods for extracting the CP-violating
parameters $\rho$ and $\eta$ of the Cabibbo-Kobayashi-Maskawa (CKM)
quarks mixing matrix, or, equivalently, $\alpha$, $\beta$ and $\gamma$
of the unitarity triangle (for a review of CP violation in the
Standard Model (SM), see Ref.~\cite{londonmrst}). However, most
of these make significant assumptions, leading to large associated
theoretical errors. Also, in the last few years, the BaBar and Belle
experiments have measured the branching ratios of rare $B$ decays with
a precision sufficient to challenge many theoretical calculations.
This is particularly true for $B \to \pi K$ decays. It is possible to
extract CP-violating parameters from these decays alone using the old
Nir and Quinn (NQ) analysis \cite{nirquinn}. I will give a brief
review of this method in Sec.~\ref{secnirquinn}.  Unfortunetly, we
know that the NQ method is incorrect since their assumption of
neglecting electroweak-penguin (EWP) contributions is clearly false.

The main purpose of this paper is to describe an extension of the NQ
analysis. I will show in Sec.~\ref{secewp} how to use SU(3) flavor
symmetry to take into account the effect of EWP's in $B \to \pi K$
decays, and thus, to resuscitate the NQ method. Finally, I will
discuss briefly the possibility of detecting and measuring new physics
(NP) in Sec.~\ref{secnp}. 

\section{Preliminaries}\label{secbas}

I begin with some preliminaries concerning $B \to \pi K$ decays. We
are interested in 8 decays: $B^0 \to \pi^0 K^0$, $B^0 \to \pi^- K^+$,
$B^+ \to \pi^0 K^+$, $B^+ \to \pi^+ K^0$, and their CP-conjugate
processes. Their measurement leads to a total of 9 experimental
quantities:
\begin{itemize}

\item 4 averaged branching ratios
\beq
\Gamma (B \to f) + \Gamma (\bar B \to {\bar f}) ~,
\eeq

\item 4 time-independent direct CP asymmetries
\beq
\frac{\Gamma (B \to f) - \Gamma (\bar B \to {\bar f})} {\Gamma (B \to
f) + \Gamma (\bar B \to {\bar f})}~,
\eeq

\item 1 time-dependent indirect CP asymmetry
\beq
\frac{\Gamma (B^0 (t) \to f) - \Gamma (\bar B^0 (t) \to {\bar f})}
{\Gamma (B^0 (t) \to f) + \Gamma (\bar B^0 (t) \to {\bar f})} = -
{\rm Im}(\lambda) \sin(\Delta M_B t)~,
\eeq
where $\lambda \equiv e^{-2 i \beta} {\bar A}/A$, and $A \equiv \Gamma
(B \to f)$, $\bar A \equiv \Gamma (\bar B \to {\bar f})$ involve the
CKM phase $\beta$.

\end{itemize}

Recent experimental measurements are presented in Table~\ref{tab1}
\cite{hfag} for completeness. Indirect CP violation has been
measured recently by BaBar: $S_{\pi^0 K_S} = 0.48^{+0.38}_{-0.47} \pm
0.06$ \cite{babar}. It is clear that uncertainties are big, and
we cannot hope for a miracle for resulting constraints on CKM
parameters.

\begin{table}[ph]
\caption{Averaged branching ratios and direct CP asymmetries.\vspace*{1pt}}
\centering
{
\begin{tabular}{|c|c|c|}
\hline
{} &{} &{}\\[-1.5ex]
{} &Branching ratio($10^{-6}$) & $A_{CP}$ \\[1ex]
\hline
{} &{} &{}\\[-1.5ex]
$B^0 \to \pi^0 K^0$ &$11.7\pm 1.4$ &$0.11\pm 0.23$ \\[1ex]
$B^0 \to \pi^- K^+$ &$18.2\pm 0.8$ &$-0.095\pm 0.028$\\[1ex]
$B^+ \to \pi^0 K^+$ &$12.5^{+1.1}_{-1.0}$ &$-0.00\pm 0.05$\\[1ex]
$B^+ \to \pi^+ K^0$ &$21.8\pm 1.4$ &$0.03\pm 0.04$\\[1ex] 
\hline
\end{tabular}\label{tab1}  }
\vspace*{-13pt}
\end{table}

Another important aspect of $B \to \pi K$ decays is the isospin
quadrilateral \cite{gliso}. Under isospin symmetry, mesons form
isospin multiplets, e.g.\ $(B^+,B^0)$, $(\pi^+,\pi^0,\pi^-)$ and
$(K^+,K^0)$. The effective hamiltonian for $B \to \pi K$ can then be
written as a linear combination of the $(I=0, I_3=0)$ and $(I=1,
I_3=0)$ isospin components. It is then easy to show, using the
Wigner-Eckart theorem and a table of Clebsch-Gordan coefficients, that
amplitudes of $B \to \pi K$ decays obey a simple quadrilateral rule in
the complex plane:
\bea 
A^{+0}+\sqrt{2} A^{0+} &=& \sqrt{2} A^{00} + A^{-+}~,\nn\\
\bar A^{+0}+\sqrt{2} \bar A^{0+} &=& \sqrt{2} \bar A^{00} + \bar
A^{-+} ~,
\label{eqquad} 
\eea 
where $A^{ij}=A(B\to \pi^i K^j)$ and $\bar A$'s are CP-conjugated
amplitudes.

In addition, we can describe the decays in term of Feynman diagrams
\cite{ghlr1}. To lowest order, there are 6 diagrams involved in
$B \to \pi K$ decays: a gluonic penguin amplitude ($P$), a
color-favored tree amplitude ($T$), a color-suppressed tree amplitude
($C$), an annihilation amplitude ($A$), a color-favored
electroweak-penguin amplitude ($\pew$) and a color-suppressed
electroweak-penguin amplitude ($\pewc$). We can easily derive the
following relations :
\bea
A^{+0} &=& P - \frac{1}{3} \pewc + A \expg ~, \nn\\
\sqrt{2} A^{0+} &=& -P - T \expg - C \expg - \pew - \frac{2}{3} \pewc
- A \expg ~,\nn\\
\sqrt{2} A^{00} &=& P - C \expg - \pew - \frac{1}{3} \pewc ~,\nn\\
A^{-+} &=& -P-T\expg -\frac{2}{3} \pewc ~,
\label{eqamplitudes}
\eea
where the weak phase $\gamma$ is written explicitly, and the strong
phases are included in the amplitude definitions for notation
convenience. For CP-conjugated amplitudes, we have exactly the same
relations as in Eq.~(\ref{eqamplitudes}), but with $e^{-i \gamma}$
instead of $\expg$. Note that isospin quadrilateral equations
(Eq.~(\ref{eqquad})) are respected by Eq.~(\ref{eqamplitudes}).

Since our main goal is to extract CKM weak phases from $B \to \pi K$
decays alone, let us ask the following question: do we have enough
information to extract CKM weak phases?  A simple counting exercise
tells us that the answer is no. We have 13 independent theoretical
parameters: 6 magnitudes ($|P|$, $|T|$,$|C|$,$|\pew|$,$|\pewc|$ and
$|A|$), 5 relative strong phases and 2 CKM weak phases. But we have
only 9 experimental measurements, which is not sufficient to solve the
full system of equations. A natural approach for such a deadlock is to
neglect some amplitudes until we can solve. This leads us to the NQ
analysis.

\section{Nir and Quinn Analysis}\label{secnirquinn}

I present here a very summarized version of the NQ method
\cite{nirquinn}. The basic assumption of NQ is to neglect all
EWP's, reducing the number of theoretical parameters to 9. The system
is now solvable in principle.

We have seen that the decay amplitudes form two isospin quadrilaterals
in the complex plane. Let us assign them the names {\it quadrilateral}
and {\it quadrilateral bar}. The first step is to rotate {\it
quadrilateral bar} by an angle of $2 \gamma$. I will use the term {\it
quadrilateral tilde} for the result:
\beq
\tilde A ^{ij} = e^{2 i \gamma} \bar A^{ij}~.
\eeq
This rotation done, there are two important observations :
\begin{enumerate}

\item The {\it quadrilateral} and the {\it quadrilateral tilde} have a
common diagonal:
\bea
D_1 &=& A^{-+} + \sqrt{2} A^{00} = (T + C) \expg~,\nn\\
\tilde D_1 &=& \tilde A^{-+} + \sqrt{2} \tilde A^{00} = e^{2 i \gamma}
(T + C) e^{- i \gamma} = D_1 ~.
\eea

\item The other two diagonals bisect one another:
\beq
A^{00} + A^{0+} = \tilde A^{00} + \tilde A^{0+}~.
\eeq

\end{enumerate}
With these two observations, there is enough information to fix the
quadrilaterals, up to a discrete ambiguity. The CKM weak phase
$\alpha$ is then extracted using the time-dependent CP asymmetry and
the angle between $Arg(\bar A^{00} / A^{00})$ taken from
quadrilaterals.  Our goal seems to be reached, but unfortunatly, this
analysis is incorrect since EWP's are not negligible. It was shown
\cite{ghlr2} using factorization that amplitudes obey the
hierarchy
\bea
O(1) &:& |P|~,\nn\\
O(\epsilon) &:& |T|, |\pew|~,\nn\\
O(\epsilon^2) &:& |C|, |\pewc|~,\nn\\
O(\epsilon^3) &:& |A|~,\label{eqhierarchy}
\eea
where $\epsilon \approx 0.2$. Even if this hierarchy is very rough, it
is clear that neglecting EWP's implies large theoretical errors. This
problem is known as {\it EWP pollution}. For this reason, the
extraction of CKM weak phases from $B \to \pi K$ decays alone was
abandoned for a decade. To solve this problem, we need more
theoretical assumptions.

\section{Taking into account EWP's}\label{secewp}

I begin this section with a quick review of the effective
hamiltonian. Using the renormalization group and operator-product
expansion, the effective hamiltonian can be written as a linear
combination of simple operators \cite{bblau}:
\beq
H_{\sss eff} = \frac{G_F}{\sqrt{2}} \sum_{q=d,s} \left(( V^*_{ub}
V_{uq}(c_1 O_1 +c_2 O_2)-V^*_{tb} V_{tq} \sum^{10}_{i=3} c_i o_i)
\right) ~,
\eeq
where
\bea
O_1 =(\bar b_{\beta} u_{\alpha})_{\sss V-A} (\bar u_{\alpha}
q_{\beta})_{\sss V-A}~, & & O_2 =(\bar b u)_{\sss V-A} (\bar u
q)_{\sss V-A}~,\nn\\
O_3 =(\bar b q)_{\sss V-A} \sum_{q'} (\bar q' q')_{\sss V-A}~, & & O_4
=(\bar b_{\beta} q_{\alpha})_{\sss V-A} \sum_{q'} (\bar q'_{\alpha}
q'_{\beta})_{\sss V-A}~,\nn\\
O_5 =(\bar b q)_{\sss V-A} \sum_{q'} (\bar q' q')_{\sss V+A}~, & & O_6
= (\bar b_{\beta} q_{\alpha})_{\sss V-A} \sum_{q'} (\bar q'_{\alpha}
q'_{\beta})_{\sss V+A}~,\nn\\
O_7 = \frac{3}{2} (\bar b q)_{\sss V-A} \sum_{q'} (\bar q' q')_{\sss
V+A}~, & & O_8 =\frac{3}{2} (\bar b_{\beta} q_{\alpha})_{\sss V-A}
\sum_{q'} (\bar q'_{\alpha} q'_{\beta})_{\sss V+A}~,\nn\\
O_9 = \frac{3}{2} (\bar b q)_{\sss V-A} \sum_{q'} (\bar q' q')_{\sss
V-A}~, & & O_{10} = \frac{3}{2} (\bar b_{\beta} q_{\alpha})_{\sss V-A}
\sum_{q'} (\bar q'_{\alpha} q'_{\beta})_{\sss V-A}~.
\eea
Above, the $c$'s are the well known Wilson coefficients, $O_1$ and
$O_2$ are tree operators, $O_3$ to $O_6$ are gluonic penguin operators
and $O_7$ to $O_{10}$ are EWP operators. There are two useful
observations to make here. First, the Wilson coefficients $c_7$ and
$c_8$ are small compared with $c_9$ and $c_{10}$.
\bea
c_7 = 3.49 \times 10^{-4}~, & c_8 = 3.72 \times 10^{-4}~,\nn\\
c_9 = -9.92 \times 10^{-3}~, & c_{10} = 2.54 \times 10^{-3}~.
\eea
The second observation is that, under the assumption of neglecting
$O_7$ and $O_8$, EWP operators are purely $(V-A)\times (V-A)$ and have
exactly the same structure as the tree operators after a Fiertz
transformation. We can therefore guess that there exist relations
relating trees and EWP's. In fact, using SU(3) flavor symmetry and
neglecting annihilation amplitude, we can derive the following
explicit relations \cite{ipll} :
\bea
\pew = \frac{3}{4} \frac{c_9+c_{10}}{c_1+c_2} R (T+C)+\frac{3}{4}
\frac{c_9-c_{10}}{c_1-c_2} R (T-C)~,\nn\\
\pewc = \frac{3}{4} \frac{c_9+c_{10}}{c_1+c_2} R (T+C) - \frac{3}{4}
\frac{c_9-c_{10}}{c_1-c_2} R (T-C)~,
\eea
where $R=|V^*_{tb} V_{ts} / V^*_{ub} V_{us}| = (\lambda^2 \sqrt{\rho^2
+ \eta ^2})^{-1}$. We have therefore written EWP's in terms of trees,
CKM factors and known quantities.

Let us repeat the counting exercise using these new relations: we have
3 magnitudes ($|P|$, $|T|$,$|C|$), 2 relative strong phases and 2 CKM
weak phases for a total of 7 independent theoretical parameters. We
have a total of 8 measurements (neglecting annihilation amplitudes
implies that $|A^{+0}|$ = $|\bar A^{+0}|$). The consequence is that,
in principle, we can solve for the full unitarity triangle with $B \to
\pi K$ decays alone. Note also that the time-dependent CP asymmetry is
not necessary. Naturally there are many discrete ambiguities, but the
measurement of the indirect CP asymmetry or other outside inputs
(e.g.\ the measurement of $\sqrt{\rho ^2 + \eta ^2}$) reduces these
significantly.

Finally, we made several assumptions to get this result. Let us
examine them one by one and analyse roughly the associated errors.
First, we have neglected anihilation contribution. According to the
hierarchy of Eq.~(\ref{eqhierarchy}), this is an error at the order of
1\%.  Second, I did not mention it previously, but there are in fact
three gluonic penguins and three EWP's (depending on the internal $u$,
$c$ or $t$ quark in the loop). We have supposed that all penguins are
dominated by the internal $t$ quark. From CKM factors, we can estimate
this error to be about 2\%. Third, we have neglected the operators
$O_7$ and $O_8$ in the effective hamiltonian making about a 4\% error.
Finally, the SU(3)-breaking is estimated to be about 5\% for $B \to
\pi K$ decays \cite{neubertrosner} (though it is typically larger
than this). Adding these estimates, we find that the total error is
roughly 10\%. With such a theoretical error, it is clear that the
extracted values of CKM phases are not clean. This method alone cannot
provide precise values of CKM parameters. In practice, this method is
applicable in parallel with other methods in a more global fit (e.g.\
see Ref.~\cite{ckmfitter}).

\section{New Physics}\label{secnp}

Some work has been done analyzing $B \to \pi K$ decays in the
framework of SM. Everything doesn't seem to be fine. As an example, there is a $2.4 \sigma$ deviation of the {\it Lipkin sum rule} and
some ratios of decay rates are not as expected \cite{grnp}.  Also,
there are some stronger signs of discrepency in other decays, such as
the $3.5 \sigma$ discrepency between the measurement of the CP
asymmetry in $B^0_d (t) \to J/\psi K_s$ and that in $B^0_d (t) \to
\phi K_s$ from Belle. All of these have in common $\bar b \to \bar s$
transitions. Thus, even if the discrepencies are not huge, the door is
open for some scenarios of new physics (NP). These are only hints for
NP, but it might be interesting to go further and to measure	
theoretical parameters in some NP scenarios.  It is possible in the $B$
system (e.g. see Ref. \cite{reck}).

In the previous section, we have not exploited the full potential of
the SU(3) flavor symmetry. In fact, under this symmetry, $B \to \pi
\pi$, $B \to \pi K$ and $B \to K K$ decays (and others involving
$\eta$'s) are related to one another \cite{gpy}. This adds many
experimental measurements, so that there are many new constraints on
the system. It is true that flavor symmetry is not a perfect symmetry,
but even if theoretical errors associated with this assumption are
large, it gives us the possibility to add many new theoretical
parameters. Thus, the $B$ system gives us enough information to test
some scenarios of NP.

It has been shown recently \cite{dattalondonnp} that, to a good
approximation, NP amplitudes have negligible strong-phase differences.
This is a consequence of the hypothesis that strong phases are due
mainly to rescattering. Since NP rescattering is estimated to be about
5\% of the NP contributions, we can neglect the NP strong phases.
Using this simple assumption, we can find a model-independent
parametrization of NP amplitudes.

As an example \cite{ailpss}, to make a minimal usage of the
previous hypothesis, we can combine $B \to \pi K$ and $B \to \pi \pi$
decays, and take $\beta$ from the time-dependent CP asymmetry from $B
\to J/ \psi K_S$. (It is assumed that NP affects only $\bar b \to \bar
s$, so that $B \to \pi \pi$ is not affected.) There are a total of 16
experimental measurements. With this, there is enough information to
solve for the NP parameters. In the case of totally model-independent
NP, there are 16 independent theoretical parameters, so that the
discrete ambiguities make this process a bit discouraging. On the
other hand, for more constrained NP scenarios (e.g.\
isospin-conserving NP or Z-mediated flavor-changing neutral currents),
the situation seems to be more realistic in practice. However, in
addition to the theoretical errors discussed in Sec.~\ref{secewp},
there is SU(3) breaking in the relation of $B \to \pi K$ and $B \to
\pi \pi$ decays, plus the assumption of neglecting the NP strong
phases. Thus, without a better understanding of SU(3)-breaking, the
resulting theoretical errors are big ($\gsim 25\%$). Still, even in
this case, the method could help us qualitatively to prefer some
scenarios of NP and maybe eliminate some if we are lucky enough. In
the next few years, with a better experimental precision of branching
ratios and CP violation, this procedure might be possible.

\section{Conclusion}\label{secconc}

There are three points to take from this talk. First, some reasonable
assumptions allow us to remove the EWP pollution in $B \to \pi K$
decays. Second, once removed, these decays alone suffice to extract
the full unitarity triangle with or without time-dependent
asymmetry. Finally, in principle, there is enough information in the
$B$ system to detect NP and also to measure its parameters in a
model-independent way.

\section*{Acknowledgments}

This work was financially supported by NSERC of Canada.

\renewcommand{\theequation}{A.\arabic{equation}}

\end{document}